\documentclass[reprint,amsfonts,showpacs,pas,aip,english,superscriptaddress,float]{revtex4-1}

\usepackage{graphicx}
\usepackage{dcolumn}
\usepackage{bm}
\usepackage{color}
\usepackage{amssymb}
\usepackage{natbib}
\usepackage{overpic}
\usepackage{times}

\begin{document}

\title{Coherent keV backscattering from plasma-wave boosted relativistic electron mirrors}

\author{F. Y. Li}
\affiliation{Key Laboratory for Laser Plasmas (Ministry of Education) and
Department of Physics and Astronomy, Shanghai Jiao Tong University, Shanghai 200240, China}

\author{Z. M. Sheng}
\email[]{zhengming.sheng@strath.ac.uk or zmsheng@sjtu.edu.cn}
\affiliation{Key Laboratory for Laser Plasmas (Ministry of
Education) and Department of Physics and Astronomy, Shanghai Jiao
Tong University, Shanghai 200240, China} \affiliation{SUPA,
Department of Physics, University of Strathclyde, Glasgow G4 0NG,
UK}

\author{M. Chen}
\email[]{minchen@sjtu.edu.cn}
\affiliation{Key Laboratory for Laser Plasmas (Ministry of Education) and
Department of Physics and Astronomy, Shanghai Jiao Tong University, Shanghai 200240, China}

\author{H. C. Wu}
\affiliation{Institute for Fusion Theory and Simulation, Zhejiang University, Hangzhou 310027, China}

\author{Y. Liu}
\affiliation{Key Laboratory for Laser Plasmas (Ministry of Education) and
Department of Physics and Astronomy, Shanghai Jiao Tong University, Shanghai 200240, China}

\author{J. Meyer-ter-Vehn}
\affiliation{Max-Planck-Institut f\"{u}r Quantenoptik, D-85748 Garching, Germany}

\author{W. B. Mori}
\affiliation{University of California, Los Angeles, California 90095-1547, USA}

\author{J. Zhang}
\affiliation{Key Laboratory for Laser Plasmas (Ministry of Education) and
Department of Physics and Astronomy, Shanghai Jiao Tong University, Shanghai 200240, China}

\begin{abstract}
A new parameter regime of laser wakefield acceleration driven by
sub-petawatt femotsecond lasers is proposed, which enables the
generation of relativistic electron mirrors further accelerated by the plasma wave.
Integrated particle-in-cell simulation including the mirror formation and
Thomson scattering demonstrates that efficient coherent backscattering up to keV
photon energy can be obtained with moderate driver laser intensities
and high density gas targets.
\end{abstract}

\maketitle

Recently tremendous interest for x-ray generation has been drawn
to intense laser-matter interactions. They promise ultrashort
ultracompact x-ray sources compared to large conventional
facilities such as x-ray free electron lasers~\cite{mcneil2010x}.
Various schemes have so far been developed including high harmonic
generation from either
gas~\cite{corkum2007attosecond,krausz2009attosecond} or solid
targets~\cite{teubner2009high}, laser-driven $K\alpha$
sources~\cite{chen2008study,chen2010intense}, plasma acceleration
based betatron radiations~\cite{corde2013femtosecond}, etc. Among
these schemes, a simple and efficient approach is based on
laser-driven relativistic electron mirrors. They may compress
femtosecond probe pulses to attoseconds and boost photon energy by
factors $\Gamma=4\gamma_x^2$, where
$\gamma_x=(1-v_x^2/c^2)^{-1/2}$ is the relativistic Lorentz factor
related to the mirror's normal velocity $v_x$. Laser-driven
plasmas provide a rich source of such mirrors
~\cite{bulanov2013relativistic} and some involving interaction with
solids~\cite{teubner2009high,esirkepov2009boosted,meyer2009coherent,wu2010uniform,kiefer2013relativistic}
are dense enough for coherent backscattering.

In underdense plasma, density crests of strongly driven plasma
waves (or when close to wave breaking) develop converging spikes
and have also been suggested as relativistic flying
mirrors~\cite{decker1996evolution,bulanov2003light}. This regime
is of particular interest for experiments because high repetition
rates are possible with gas targets as well as using less
challenge conditions compared to the solid schemes which demand
extremely high laser contrast. However, there remains a couple of
issues when using the flying mirror for coherent backscattering up
to keV photon energy. They are mainly due to the limitation for
the Doppler factors determined by the phase velocity of the plasma
wave ($v_p$), i.e.,
$\Gamma\leq4\gamma_p^2=4/(1-v_p^2/c^2)\simeq4n_c/n_0$; $n_0$ and
$n_c$ are the ambient electron density and the critical density,
respectively~\cite{tajima1979laser,bulanov2003light}. The existing
experiments
~\cite{kando2007demonstration,kando2009enhancement,pirozhkov2007frequency}
have suggested limited $\Gamma<100$ for $n_0$ at a few
$10^{19}~\rm cm^{-3}$. Reducing $n_0$ can give larger $\Gamma$,
but also requiring stronger driver to approach the wavebreaking
limit~\cite{benedetti2013numerical}. Meanwhile, it is preferred
that a broad laser focal spot, e.g.,
$\sigma_0>\sqrt{a_0}\lambda_p/\pi$, is used to drive the wake in
the quasi-one-dimensional (quasi-1D) regime
~\cite{matlis2006snapshots,li2014radially} so that it can provide
a flat mirror plane for collimated backscattering. Here,
$\lambda_p$ is the plasma wavelength and
$a_0=8.5\times10^{-19}\rm\lambda_0[\mu m]\sqrt{I_0[W/cm^2]}$ is
the normalized laser amplitude with $\lambda_0$ the laser
wavelength and $I_0$ the peak intensity. Tremendous incident peak
power, $P_0\propto a_0^2\sigma_0^2\propto n_0^{-4}$, is then
required at low $n_0$ for high $\Gamma$ factors. More critically,
the exact breaking point, crucial for efficient backscattering, is
actually hard to achieve in quasi-1D thermal
plasmas~\cite{solodov2006limits}. Instead, wave breaking often
occurs as the laser pulse evolves significantly (including
self-modulation and self-focusing) and drives the wake in the
bubble regime~\cite{pukhov2002laser,lu2007generating}
with an evidence of generating
energetic electron beams~\cite{pirozhkov2007frequency}. As a
result, backscattering off such near-spherical bubble shell leads
to extremely large radiation divergence.

The aim of this Letter is to report a scenario that can overcome
the above shortages and result in coherent Thomson backscattering
up to keV photon energy using reasonable driver conditions. The
key point is to break the constraint set by the wave's phase
velocity $v_p$. The way to achieve this is to drive the wake even
harder and let the mirror be properly injected into the plasma
wave. Consequently, the Doppler factor of the mirror is boosted by
wakefield acceleration~\cite{esarey2009physics}, not limited any
more by $v_p$. As we shall see, the only limitation now turns out
to be $\gamma_x$ rising above some threshold, at which the
scattered light degrades into incoherent pulses. With this
scenario low density $n_0$ is not necessarily used for high
$\Gamma$ factors. Instead, high density gas targets can be employed
which, as we have argued, will reduce the peak laser power
required.

It is clear that synchronous injection of the singular density
crest is desired so that the mirror spike can be accelerated as a
whole. Normally, quasi-continuous injection is found, which
produces femtosecond narrow electron bunches. Such bunches have
been used for incoherent
backscattering~\cite{schwoerer2006thomson,phuoc2012all,xu2014spectrum}.
To make the sharp injection possible, two steps are necessary. The
first is to drive the wake close to breaking but without
injection. Specifically, a density upramp with proper gradients
can be used to suppress electron injection for the first few wave
buckets behind the driver~\cite{li2013dense,mu2013robust}. It is
due to the fact that, along the ramp, the wave can travel at a
superluminal phase speed even for high nonlinearity. As a result,
the density crests can be stably compressed into spikes without
premature injection. The free of injection eventually terminates
as the laser pulse propagates through the ramp to a following uniform
density where the wave's phase speed becomes subluminal.
This refers to the second step as sharp injection.
As the phase speed falls below the light speed abruptly at the
density transition region, a major part of the tightly compressed
density crest is injected as a whole. Details of the controlled
injection are described in a previous publication~\cite{li2013dense}.

This sharp injection actually works over a wide range of plasma
densities, and for given density the laser amplitude only
has to meet some threshold. Here, in order to drive the boosted
mirrors with reasonable laser conditions, we propose using high density
gas plasmas ($\sim10^{20}~\rm cm^{-3}$) so that a laser focal spot
of $10\sim20~\rm\mu m$ is sufficient for wake excitation in the
quasi-1D regime. Notice that most experiments on wakefield
acceleration are currently operated in the bubble regime in low
densities ($10^{17}\sim10^{18}~\rm cm^{-3}$) for generating GeV
beams~\cite{kim2013enhancement,wang2013quasi}. However, it is the
essence of the new parameter regime that allows for the trapping
of dense electron sheets and the consequent generation of coherent
keV backscattering instead of incoherent sources normally obtained
so far~\cite{corde2013femtosecond,schwoerer2006thomson,phuoc2012all}.

\begin{figure}[t]
\centering
\includegraphics[width=0.34\textwidth]{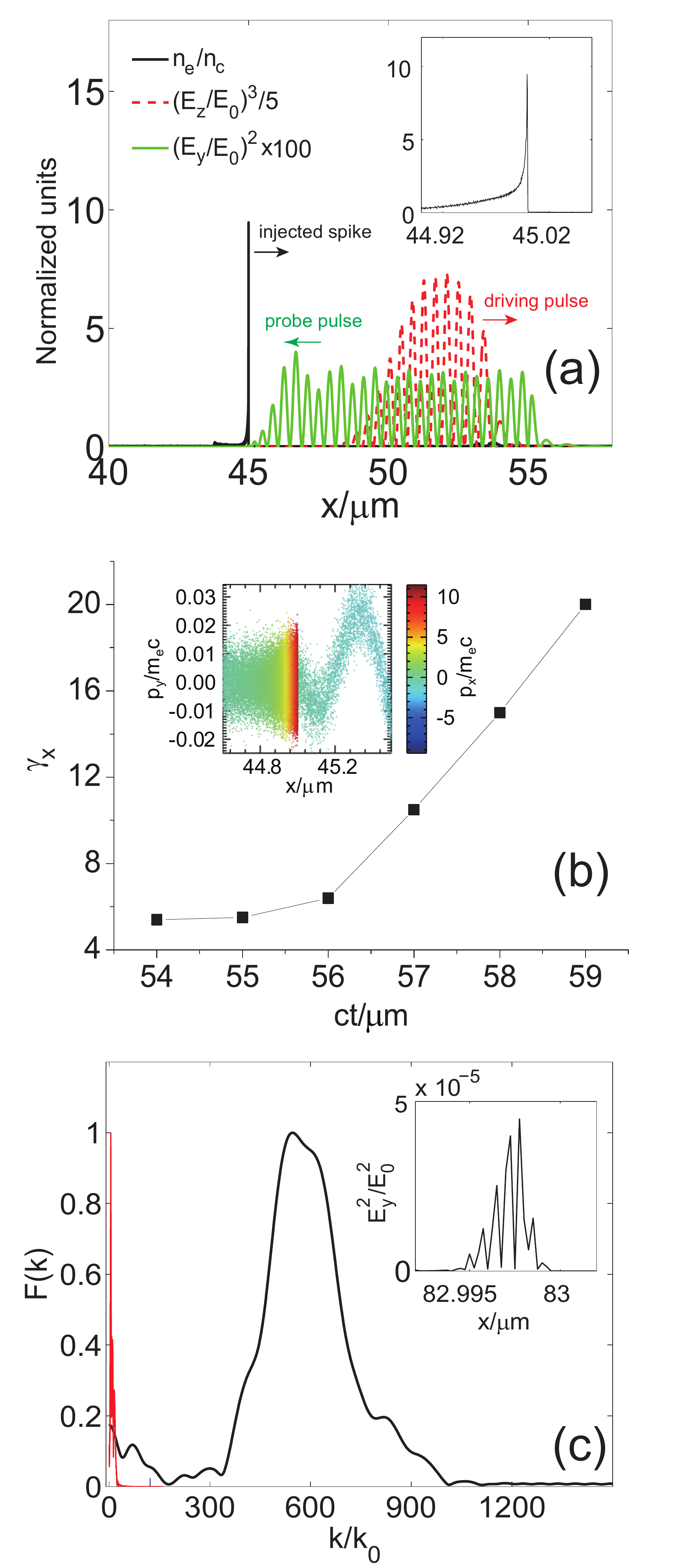}
\caption{(color online) Results of 1D PIC simulation. (a) Snapshot
of electron density $n_e$, driving pulse $E_z^2$, and probe pulse
$E_y^2$ at $t=57~\rm\mu m/c$ shortly after the first density spike
is injected; inset shows a closer look of the spike profile with
corresponding phase space $x$-$p_y$ (colored according to $p_x$)
in the inset of (b). (b) Evolution of peak $\gamma_x$ of the
mirror spike around the injection instant; (c) Spatial spectrum of
the scattered pulse $E_y$ (corresponding intensity profile $E_y^2$
in the inset) from the injected spike. The red curve refers to the
spectrum obtained for the driving laser $a_0=2.5$ while keeping
the other parameters fixed.}
\end{figure}

Below we demonstrate coherent backscattering off the injected mirror
via particle-in-cell (PIC) simulations~\cite{fonseca2002lect}.
At first 1D PIC simulations are used to illustrate the basic features.
The interaction geometry is shown in Fig. 1(a). A 12.5-cycle driving laser,
polarized along $z$ axis, propagates in the $+x$ direction.
Its dimensionless amplitude is $a_0=6$, corresponding to
$I_0=7.7\times10^{19}~\rm W/cm^2$ for $\lambda_0=0.8~\rm\mu m$.
The plasma density is $n_0=7\times10^{19}~\rm cm^{-3}$
at $x\in[45, 75]~\rm\mu m$ with a $45~\rm\mu m$ long ramping front.
The probe pulse, polarized along $y$ axis,
propagates in the opposite direction and is appropriately delayed
so that it meets the first density spike shortly after the sharp injection,
i.e., the moment depicted in Fig. 1(a). The probe pulse
with amplitude $a_{pr}=0.1$ has the same frequency as the driver
($\omega_{pr}=\omega_0=2\pi c/\lambda_0$) and takes a 12.5-cycle
rectangle shape in time domain. In the simulation,
4000 cells per micron are used to resolve the high-frequency backscattering.
An isotropic electron temperature is initialized, e.g., 10s eV
in the transverse direction and much lower in the longitudinal direction.
The low longitudinal temperature is acceptable in experiments
as the electrons released from atoms by prepulse ionization are
first dominant in transverse quiver motion at subpicosecond time scale.

The inset of Fig. 1(a) shows a closer look of the injected density
spike. A remarkable feature is the very sharp front
edge~\cite{bulanov2005spectral} which, as expanded in phase space
$x$-$p_y$ [see the inset of Fig. 1(b)], shows a monoenergetic peak
of $p_x\sim10$ or $\gamma_x\simeq10.5$. The evolution of this peak
$\gamma_x$ around the injection instant is given in Fig. 1(b).
It is seen that $\gamma_x$ is boosted by wakefield acceleration after
$t=56~\rm\mu m/c$. Before that it is kept at about 5.5, in
consistent with the estimation from
$\gamma_p\simeq\sqrt{n_c/n_0}=5$. Fig. 1(c), most importantly,
presents the spectrum of the scattered pulse.
It exhibits an ultrabroad bandwidth extending up to
$k_x/k_0\simeq1000$ or 1.5 keV in photon energy, with
$k_0=2\pi/\lambda_0$. The corresponding spatial profile (see the
inset therein) consists of only 3.5 cycles, rather than 12.5
cycles for the probe pulse. This self-contraction
effect~\cite{wu2011nonlinear} is mainly ascribed to high enough
$\gamma_x$, for which the coherent reflection
condition~\cite{wu2009reflectivity} (i.e.,
$n_e\gg10^{13}\gamma_x^4~\rm cm^{-3}$) is no longer satisfied and
the backscattered light becomes incoherent and orders of magnitude
weaker in the intensity. That means the scattered pulse can
adjust itself as a few-cycle attosecond x-ray pulse regardless
of the probe pulse length. Here, the coherent backscattering
terminates at about $t=58.5~\rm\mu m/c$ related to
$\gamma_x\simeq16$ or $\Gamma=1024$, which is in fair agreement
with the maximal upshifted frequency observed. For comparison the
spectrum of reflection from non-breaking density crest is also
plotted in Fig. 1(c); only $\Gamma\simeq16$ is obtained
corresponding to $\gamma_x\simeq2$.

The maximal amplitude of the scattered pulse
[see the inset of Fig. 1(c)] is $E_{s,peak}/E_0=0.0063$ or $25~\rm GV/m$,
corresponding to a reflectivity of $E_{s,peak}/E_{pr}=6.3\%$;
$E_0=m_e\omega_0c/e$ is the normalizing field and $E_{pr}=a_{pr}E_0$
is the probe pulse field strength. The coherent reflectivity can also
be estimated using the basic 1D model~\cite{wu2009reflectivity}.
As implied in Fig. 1(a), the mirror spike can be approximated by
$n_s(x)=n_{s0}\exp(x/d)$ for $x\leq0$ (whereas zero for $x>0$);
$d$ is a characteristic thickness. The backscattering amplitude in
the normal direction is then derived as
\begin{equation}
        \label{reflectivity}
\eta=C\pi\frac{\Omega_{s0}}{\omega_{pr}}\frac{d}{\Lambda_{s0}}
\frac{\gamma_x^2(1+\beta_x)}{\gamma},
\end{equation}
where $C=[1+16d^2\pi^2\gamma_x^4(1+\beta_x)^2/\lambda_{pr}^2]^{-1/2}$
with $\lambda_{pr}$ the probe laser wavelength,
$\Omega_{s0}=\sqrt{e^2n_{s0}/\varepsilon_0m_e}$ is the plasma frequency
associated with the maximal mirror density and $\Lambda_{s0}=2\pi c/\Omega_{s0}$.
For the simulated parameters $n_{s0}\simeq7.8n_c$, $d\simeq0.0025\lambda_0$
and $\gamma_x\simeq14$, the reflectivity amounts to $\eta\simeq14.3\%$,
nearly twice the value of the above observation.
During interaction with the mirror, the probe pulse front has already
somewhat damped after propagation in plasmas [see Fig. 1(a)];
this may explain the slight overestimation provided by the 1D model.

\begin{figure}[t]
\centering
\includegraphics[width=0.48\textwidth]{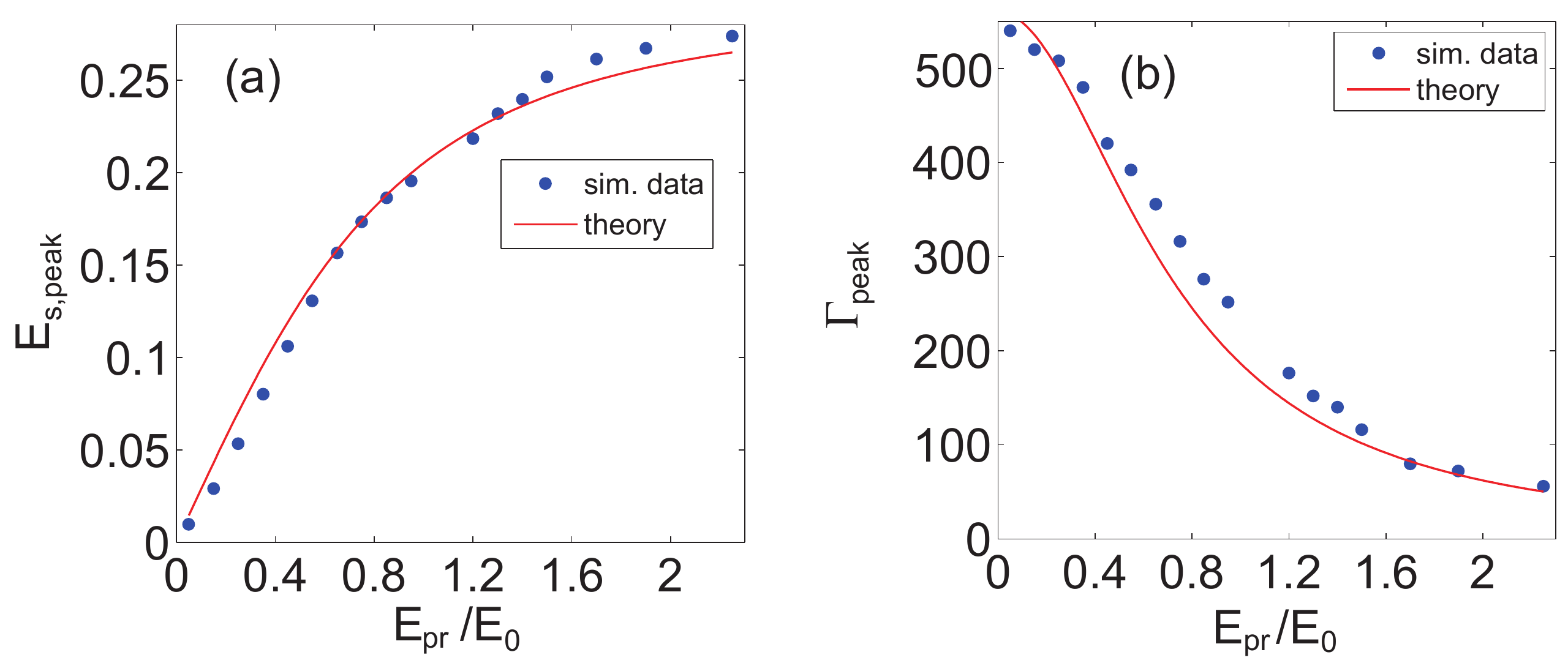}
\caption{(color online) (a) Peak scattering fields
$E_{s, peak}$ and (b) frequency upshifting factor for the
corresponding spectra peak $\Gamma_{peak}$ versus different probe
amplitudes. Blue dots and solid curves represent
1D simulation results and nonlinear theory predications, respectively.}
\end{figure}

A series of simulations with different probe laser amplitudes (up
to $E_{pr}/E_0>1$) is also conducted to study nonlinear
backscattering off the boosted mirror. Figure 2 presents the
maximal scattering fields $E_{s, peak}$ and the upshifting factors
$\Gamma_{peak}$ for the corresponding spectra peak.
For high-amplitude probe laser, its nonlinear ponderomotive force
damps the mirror energy or the Doppler factor continuously as
$E_{pr}/E_0$ increases and $E_{s, peak}$ also becomes saturated.
These nonlinear features have been derived as
$\Gamma_{peak}\simeq4\gamma_x^2/(1+E_{pr}^2/E_0^2)$ and
$E_{s, peak}=E_{sat}(E_{pr}/E_0)(1+E_{pr}^2/E_0^2)^{-1/2}$ with $E_{sat}$
the saturated amplitude~\cite{wu2011nonlinear}, and they show fair
agreement with the simulation results.

\begin{figure}[t]
\centering
\includegraphics[width=0.48\textwidth]{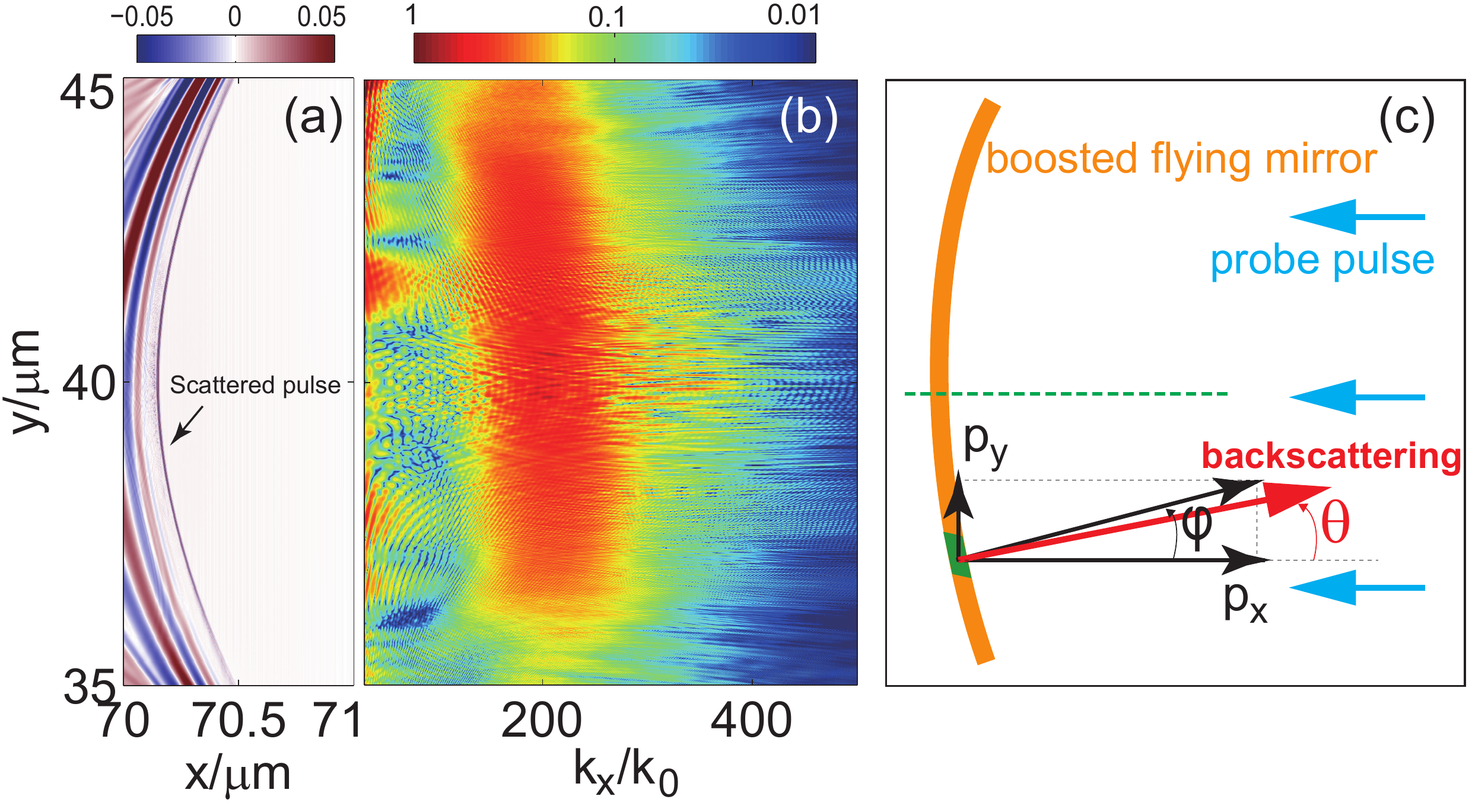}
\caption{(color online) Results of 2D PIC simulation.
(a) Spatial profile of scattered pulse $E_z/E_0$
(the probe laser is now polarized in $z$ direction while
the driver in $y$ direction) and (b) corresponding spectrum in space $k_x$-$y$.
(c) Schematic drawing of backscattering off the
boosted flying mirror, where $\varphi$ is the angle
between $\vec{p}_x$ and the total momentum $\vec{p}$,
and $\theta$ is the angle between the normal direction
of the mirror segment and the $+x$ direction.}
\end{figure}

To further explore the multidimensional effects of the boosted
flying mirror concept, high resolution 2D PIC simulations (e.g.,
with the spatial mesh grid $1800\times100$ cells per square
microns) are conducted. Notice that this resolution, though
already high, is still not sufficient to resolve the highest
frequency that could be observed in the above 1D simulations which
employed even higher resolution. To compensate this incapacity due
to limited computational resources, highly nonlinear
backscattering at $E_{pr}/E_0=1$ [the nonlinearity can reduce the
effective $\Gamma$ factors as shown in Fig. 2(b)] is performed.
The simulation parameters are similar to that for Fig. 1 except
that a focal spot of $17~\rm\mu m$ (about 1.6 times the nonlinear
plasma wavelength $\lambda_{np}$) is used to drive the wake in the
quasi-1D regime. Figure 3(a) shows the scattered pulse within a
diameter of $10~\rm\mu m$. The scattered pulse amplitude is
$E_{scat}/E_0\simeq0.1$, corresponding to a peak intensity
$\sim10^{16}~\rm W/cm^2$. The paraboloidal shape of the injected
sheet arising from nonlinear plasma frequency
shifts~\cite{matlis2006snapshots} directly maps into the
scattered pulse with a small curvature of $\sim1/40~\rm\mu
m^{-1}$, as shown in Fig. 3(a). More flat mirror planes for more
collimated x-ray radiations can be expected if a super-Gaussian
driver is used.

Additional multidimensional effects come from the transverse momentum
(e.g., $p_y$ in the present 2D geometry) of the injected mirror,
which holds even for non-breaking density wave crest
and grows continuously during wakefield acceleration
after injection~\cite{li2014radially}.
As documented in a number of papers
~\cite{wu2009reflectivity,meyer2009coherent,wu2010uniform},
the transverse momenta may make $\Gamma=4\gamma_x^2$
significantly smaller than the full Doppler factor $4\gamma^2$
with $\gamma=\gamma_x[1+(p_y/m_ec)^2]^{1/2}$.
However, the scattered spectrum shown in space $k_x$-$y$
[see Fig. 3(b)] is almost uniform transversely.
To account for this feature, we choose a segment of the thin mirror
[see the schematic drawing in Fig. 3(c)] and analyze its behavior for backscattering.
For coherent backscattering off a relativistic mirror ($\gamma\gg1$),
the scattered pulse is always directed close to the normal direction
($\vec{n}$) of the segment with the angle $\theta$ relative to the
$+x$ direction~\cite{wu2009reflectivity}. For the present quasi-1D
wave $\theta$ is approximated by $\tan\theta=-d\lambda_{np}/dy$
with $\lambda_{np}$ the plasma wavelength at each
transverse position $y$. On the other hand, the segment electrons
move at an angle defined as $\tan\varphi=p_y/p_x$.
The momentum along the normal direction is then given by
$p_n=p_x\cos\theta+p_y\sin\theta$.
The near uniform spectrum shown above requires
\begin{equation}
    \label{2}
\Gamma_n\cos\theta\geq4\gamma_x^2,
\end{equation}
where $\Gamma_n=4/(1-p_n^2/m_e^2c^2\gamma^2)$ is the relativistic Doppler factor
of this segment. With substitution of the above definitions,
Eq.~(\ref{2}) can be rewritten as
$|d\lambda_{np}/dr|\leq2p_xp_y/(p_x^2-p_y^2/2-1/2)$.
This inequality sets an upper limit (about several times the angle
of $\varphi$) for $\theta$ and can be readily met
for the small curvature mirror driven in the quasi-1D regime.
This effect can also be simply explained as that the tilt mirror
surface deflects the coherent scattering
close to the electron momenta direction, so that the
full Doppler factors are almost recovered.

The uniform spectrum shows a peak at $k_x/k_0\sim200$ which follows
precisely the nonlinear 1D results of Fig. 2(b).
This strongly suggests that the above 1D scaling for higher
upshifting factors (e.g., up to keV level)
also applies to high dimensions,
though direct verifications with 2D simulations are
restricted by higher resolutions available.
For the present case, the scattered pulse ($0.1~\rm TW$ peak power
and $4~\rm nm$ center wavelength) is able
to deliver over $10^{10}$ photons in 30 attoseconds.
The energy conversion efficiency from the probe laser to the
scattered x-ray pulse is about a few $10^{-4}$.
The remarkable feature of the high-flux
coherent x-ray sources is that they are provided by a table-top
facility typical for laser wakefield
acceleration~\cite{schwoerer2006thomson,phuoc2012all}.
They can be competitive with the large and expensive x-ray free
electron lasers in the peak power and also possess
much shorter durations. In addition, the present scenario works
in the quasi-1D regime so that it can be scaled to larger driving
focal spots delivered by petawatt lasers presently available.

In conclusion, we have proposed a new parameter regime of laser
wakefield acceleration for coherent Thomson backscattering. This
specific regime allows thin disk-like density wave crests to be
compressed and trapped in the wake wave. They may serve as boosted
mirrors with the Doppler factor no longer limited by the wave's
phase velocity. Laser-like attosecond backscatterings up to keV
photon energy have been demonstrated using PIC simulations. Since
high density gas targets are used, some hundred-terawatt lasers
focused at about $10~\rm\mu m$ are sufficient to drive the boosted
mirrors. These parameters are all within current technologies. The
experimental implementation of the present scheme would also benefit
from the advanced techniques developed for monitoring the
backscattering at femtosecond time
scales~\cite{pirozhkov2007frequency,kiefer2013relativistic}.

ZMS thanks the OSIRIS Consortium at UCLA and IST for providing
access to OSIRIS 2.0 framework. This work was supported in part by
the National Basic Research Program of China (Grant No.
2013CBA01504), the National Science Foundation of China (Grants
No. 11121504, 11374210, and 11374209), and the MOST international
collaboration project (Grant No. 2014DFG02330). Simulations were
supported by Shanghai Supercompuer Center and the center for high
performance computing at SJTU.


\end{document}